# The second knee observed in the local muon density spectra at various zenith angles

R.P. Kokoulin, M.B. Amelchakov, N.S. Barbashina, A.G. Bogdanov, D.V. Chernov, L.I. Dushkin, S.S. Khokhlov, V.A. Khomyakov, V.V. Kindin, K.G. Kompaniets, A.A. Petrukhin, V.V. Shutenko, I.I. Yashin, E.A. Yurina

*National Research Nuclear University MEPhI (Moscow Engineering Physics Institute), Moscow, 115409, Russia*

G. Mannocchi, G. Trinchero

*Osservatorio Astrofisico di Torino – INAF, Italy*

O. Saavedra

*Dipartimento di Fisica dell' Universita di Torino, Italy*

Local muon density spectra (LMDS) at various zenith angles have been reconstructed from the data of two detectors of the Experimental complex NEVOD. The inclined muon bundles at the ground level were detected with the coordinate detector DECOR, and for the near-vertical direction with the calibration telescope system (CTS) of the Cherenkov water detector. In comparison with the earlier DECOR results, the experimental statistics has been increased by 2-3 times for different ranges of zenith angle and muon bundle multiplicity and is now based on about 40,000 h of the setup operation. The live time of measurements with CTS is about 12,000 h. It is found with both setups that the slope of LMDS is increasing above the primary energy of about $10^{17}$ eV. The details of the experiment and data analysis are presented.

## 1. INTRODUCTION

EAS muon component investigation is a key for understanding of primary cosmic ray composition, energy spectrum and cosmic ray particle interaction with air nuclei in the Earth atmosphere at very high energies. Usually, muon detectors were constructed as additional parts of large-scale EAS setups and had limited capabilities to study muon component characteristics. Therefore, reconstruction of the energy spectrum of primary particles [1] was performed mainly using experimental data on the electromagnetic component of EAS. The first indirect indications on the presence of the second "knee" in the EAS spectrum were found in 1990-ies: the estimates of the slope of the spectrum measured with several giant EAS arrays around $10^{18}$ eV appeared higher than the values obtained with the compact setups for primary energies below $10^{17}$ eV (see e.g. [2]). The existence of the second knee in EAS electron size spectrum was finally established only in the 21st century due to the results obtained with several large-scale setups covering the corresponding energy range, such as KASCADE-Grande [3], Tunka-133 [4] and IceTop [5]. The first indication for the second knee existence in the spectrum of EAS muon component was obtained in 2008 [6] with the coordinate-tracking detector DECOR [7] which operates as a part of the Experimental complex NEVOD (MEPhI, Moscow) [8]. In this complex, there are two setups which may be used for investigations of the EAS muons in a wide energy range of primary particles. These are DECOR and the calibration telescope system (CTS) [9] of the Cherenkov water detector (CWD). The article presents the results of long-term measurements of the local muon density spectra (LMDS) at very high energies of primary particles obtained with the two setups at various zenith angles.

## 2. THE EXPERIMENTAL COMPLEX

The basis of the Experimental complex NEVOD (figure 1) is the Cherenkov water detector which is located at the ground level in the MEPhI campus. The complex includes also the coordinate detector DECOR and the CTS which provide the possibility of investigations of the EAS muon component in a wide energy range of primary particles ($10^{15} – 10^{18}$ eV). Eight DECOR supermodules (SMs) are arranged in the building galleries around the CWD water pool as shown in figure 1. Each supermodule consists of eight vertical planes (8.4 m$^2$) of streamer tube chambers. The angular accuracy for reconstructed muon tracks in the SM is better than 0.8°.

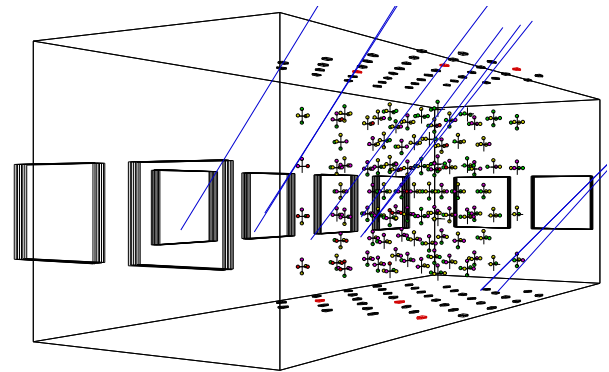

Figure 1. An example of geometric reconstruction of the event with the muon bundle in the Experimental complex NEVOD. Lines are muon tracks reconstructed from DECOR data. Small circles are hit PMTs in the CWD. Small rectangles are scintillation counters in CTS planes. Hit scintillation counters are coloured in red

The CTS consists of two planes of scintillation counters (see figure 1). In each plane, 40 counters (40×20×2 cm$^3$)

are arranged in a chess order within the area of $8\times10$ m$^2$. The top and bottom planes are located above and below the CWD volume, respectively. Primarily, CTS was constructed for calibration of quasi-spherical modules in the CWD by Cherenkov light from muon tracks [10], but after modernization, since 2013, it is also used as a detector of electron and muon EAS components.

The common triggering system of the experimental complex provides registration of multi-particle events simultaneously or separately with each setup. An example of multi-muon event is presented in figure 1.

## 3. REGISTRATION OF MUON BUNDLES

We use the local muon density spectrum (LMDS) approach [6, 9] as an instrument for investigations of high energy cosmic rays with DECOR and CTS setups.

### 3.1. Muon bundles at large zenith angles

The experiment on muon bundle registration with the DECOR detector was performed in two periods: 2002 – 2007 and 2012 – 2016. The results obtained during the first period of measurements pointed to the presence of the second knee in the EAS muon component [6] near the $10^{17}$ eV energy of primary cosmic rays. However, this indication was not statistically significant; the estimates of the LMDS slope for energies above and below this energy differed by $\Delta\beta=0.20 \pm 0.09$, i.e. a bit more than by $2\sigma$.

The procedure of muon bundle selection in DECOR data and reconstruction of the local muon density spectra from the distributions of muon bundle characteristics was described in detail earlier [11]. Differential LMDS were restored from the data on muon bundles, which were recorded in three zenith-angular ranges: 55°–60°, 60°–69°, and 69°–75° (see table 1). About 39 thousand events with muon multiplicity not less than 5 have been selected from the data accumulated for the whole observation period. The reconstructed local muon density spectra for the three pointed angular bands with effective values of zenith angles of 57°, 64° and 72° are shown in figure 2. The same local muon density at different zenith angles corresponds to different distributions of the energy of primary particles, and by increasing the zenith angle we can explore the range of higher primary energies. Arrows in the figure indicate the positions of the energy $10^{17}$ eV at different angles. These effective energy values were obtained on the basis of EAS muon component simulations with the CORSIKA program [12].

Table 1. Results of muon bundle selecting.

| Zenith angle θ, degrees | Multiplicity range | Live time, h | No. events |
|---|---|---|---|
| 55 – 60 | 5 – 9 | 21287 | 15206 |
| 55 – 60 | ≥ 10 | 21287 | 2981 |
| 60 – 69 | 5 – 9 | 31389 | 14080 |
| 60 – 69 | ≥ 10 | 41209 | 3460 |
| 69 – 75 | 5 – 9 | 31389 | 2394 |
| 69 – 75 | ≥ 10 | 41209 | 526 |

Two distinct spectrum slopes with a kink around $10^{17}$ eV primary energy are clearly seen in the figure. The integral exponents β of these LMDS (see table 2) were obtained by fitting the experimental data in the intervals corresponding to two energy ranges of primary particles: $10^{16} - 10^{17}$ eV, and more than $10^{17}$ eV. The combined estimates of the LMDS slopes in the two energy ranges derived from statistically independent results for the different zenith angles (57° and 64° for lower energies and 64° and 72° for higher energies) are presented in the last line of the table.

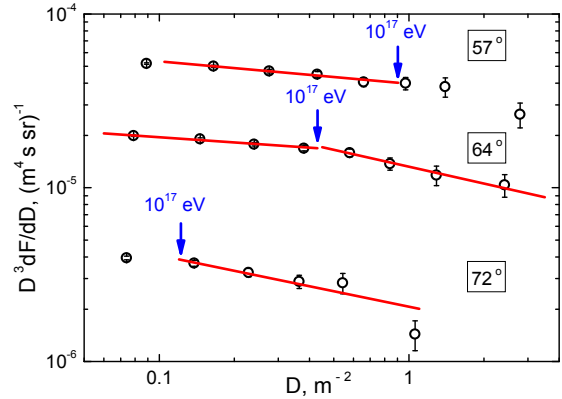

Figure 2. Differential LMDS for three values of zenith angles: 57°, 64° and 72°. Straight lines are power law fits. The arrows point to the characteristic value of the energy of primary particles $10^{17}$ eV

Table 2. Results of fitting of LMDS at 3 different zenith angles with a piece-wise power function in two ranges of effective primary energy (integral spectrum slopes β).

| Zenith angle | $E_0 = 10^{16} - 10^{17}$ eV | $E_0 > 10^{17}$ eV |
|---|---|---|
| 57° | 2.129±0.027 | - |
| 64° | 2.099±0.016 | 2.323±0.103 |
| 72° | - | 2.296±0.064 |
| Combined | 2.107±0.014 | 2.303±0.054 |

Thus, the results of the analysis of combined DECOR data on muon bundles for two observation periods confirm the presence of the second knee in EAS muon component ($\Delta\beta=0.196 \pm 0.056$) at the level of 3.5 σ.

### 3.2. Multi-muon events near the vertical direction

Investigations of multi-muon events with the CTS were performed from 2013 to 2015. Only bottom CTS plane was used for registration of EAS muon component because it is located under the 8.6 m layer of water. Events were recorded when there were at least 2 hit counters in the plane. Full live time of observations amounted to ~12,000 hours. The dependence of the event counting rate on the number of hit counters is shown in figure 3. Differential LMDS for the near-vertical direction was restored from the hit counter multiplicity distribution (~ 113,000 events containing 3 to 40 hit counters).

### 3.3. Comparison of measurement results of two setups

Differential LMDS obtained with DECOR and CTS may be directly compared for close values of the zenith angles. For this purpose, we used DECOR data on muon bundles in the zenith angels range from 30° to 40° [11] with average angle 35°; the CTS data considered here were extrapolated to the same zenith angle of 35°. The results of the comparison are shown in figure 5. Solid curves in the figure are results of calculations of the expected LMDS for two extreme assumptions on the primary composition: protons and iron nuclei. As a model of the cosmic ray energy spectrum in these calculations, a piecewise power function with the knee at the energy of 4 PeV was used. The arrows in the top of the figure indicate typical (mean logarithmic) energies of primary particles contributing to events with certain muon densities. As seen from the figure, the local muon density spectra reconstructed from the data of the two setups agree well with each other in the overlapping density range.

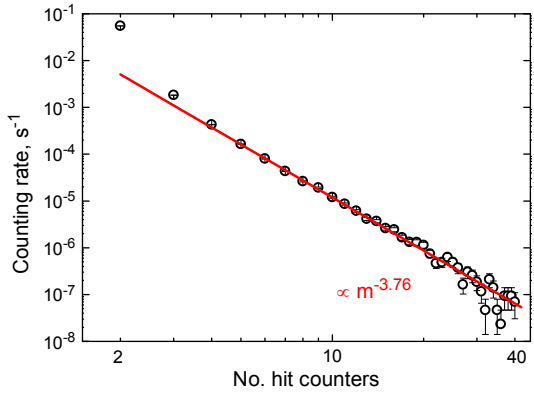

Figure 3. Dependence of the counting rate on coincidence multiplicity. Line represents a power law fit

In reconstruction of the LMDS from the CTS data, the correction for the shift of the muon density estimates due to accompanying particles was introduced. The influence of penetrating hadron EAS component and electromagnetic cascades produced in the building constructions and water on the density estimation was calculated by means of Geant4 [13] simulations of CTS plane response. According to calculations, the presence of the building and water leads to a systematic overestimation of the local muon density in the bottom CTS plane by a factor of about 1.3. The effective angle of muon bundle registration in CTS is close to 29° if we assume the angular distribution of muon bundles $\propto \cos^{4.5}\theta$ as it was measured with the DECOR detector [11]. Similar to the preceding section, the exponents of the LMDS were estimated by fitting the data (see figure 4) in two intervals corresponding to the two energy ranges of primary particles: $10^{16} - 10^{17}$ eV and more than $10^{17}$ eV. As it is seen from the figure, the local muon density spectrum at high energies becomes steeper. The change in the LMDS exponent around $10^{17}$ eV is $\Delta\beta = 0.25 \pm 0.12$ (about 2.1 $\sigma$ level). Thus, results of the measurements of the shape of the local muon density spectrum with the calibration telescope system are in a good qualitative and quantitative agreement with studies of inclined muon bundles performed with DECOR.

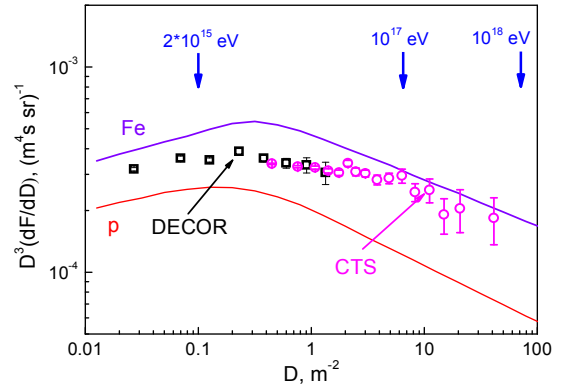

Figure 5. Differential LMDS restored from DECOR data for zenith angle 35° and CTS data extrapolated to the same zenith angle

### 4. CONCLUSIONS

The use of the LMDS approach provides the possibility of investigations of primary cosmic rays in a wide energy range ($10^{15} - 10^{18}$ eV) with relatively small size setups located at the Earth surface.

The results of the measurements of the local muon density spectra with the two setups (the coordinate-tracking detector DECOR and the calibration telescope system) of the Experimental complex NEVOD show the increase of the spectrum exponent (by $\Delta\beta \sim 0.2$) near the primary energy $\sim 10^{17}$ eV.

### Acknowledgments

Investigations were carried out in the Scientific and Educational Center NEVOD with the support of the RF Ministry of Education and Science (contract RFMEFI59114X0002, MEPhI Academic Excellence Project 02.a03.21.0005 and the government task).

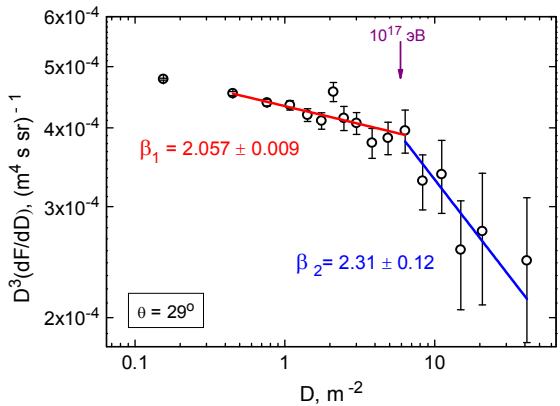

Figure 4. Differential LMDS restored from the CTS data. Straight lines are power law fits. The arrow points to the characteristic value of primary energy $10^{17}$ eV